\documentclass[a4paper,12pt, epsfig]{article}
\usepackage{epsfig}
\usepackage{amssymb,latexsym}
\usepackage{amsfonts}
\usepackage{amsmath}
\usepackage[colorlinks, linkcolor=blue, citecolor=blue, urlcolor=blue]{hyperref}
\usepackage{epstopdf}
\usepackage{graphicx}
\pdfoutput=1

\newskip\humongous \humongous=0pt plus 1000pt minus 1000pt

\newif\ifdtup

\catcode`\@=11

\@addtoreset{equation}{section}
\def\theequation{\thesection.\arabic{equation}}

\def\@normalsize{\@setsize\normalsize{15pt}\xiipt\@xiipt
\abovedisplayskip 14pt plus3pt minus3pt%
\belowdisplayskip \abovedisplayskip
\abovedisplayshortskip \z@ plus3pt%
\belowdisplayshortskip 7pt plus3.5pt minus0pt}

\def\small{\@setsize\small{13.6pt}\xipt\@xipt
\abovedisplayskip 13pt plus3pt minus3pt%
\belowdisplayskip \abovedisplayskip
\abovedisplayshortskip \z@ plus3pt%
\belowdisplayshortskip 7pt plus3.5pt minus0pt
\def\@listi{\parsep 4.5pt plus 2pt minus 1pt
      \itemsep \parsep
      \topsep 9pt plus 3pt minus 3pt}}

\relax

\catcode`@=12

\catcode`\@=11

\def\section{\@startsection{section}{1}{\z@}{3.5ex plus 1ex minus
    .2ex}{2.3ex plus .2ex}{\large\bf}}

\def\thesection{\arabic{section}}
\def\thesubsection{\arabic{section}.\arabic{subsection}}

\def\appendix{\setcounter{section}{0}
  \def\thesection{Appendix \Alph{section}}
  \def\thesubsection{\Alph{section}.\arabic{subsection}}
  \def\theequation{\Alph{section}.\arabic{equation}}}

\def\SymBoxes#1#2#3#4{\newdimen\un@t \un@t#3%
\raisebox{#1}{\rule{#2\un@t}{#4}\hskip-#2\un@t
\@tempdimb\un@t \advance\@tempdimb by-#4\@tempcntb#2\relax%
\@whilenum{\@tempcntb>0}\do{
\rule{#4}{\un@t}\hskip\@tempdimb \advance\@tempcntb by\m@ne}%
\hskip-#2\un@t \rule[\un@t]{#2\un@t}{#4}%
\rule[\un@t]{#4}{#4}\hskip-#4
\rule{#4}{\un@t}}\hskip-#4}                

\newcommand{\beq}{\begin{equation}}
\newcommand{\eeq}{\end{equation}}
\newcommand{\hsp}{,\hspace{.4cm}}
\newcommand{\bea}{\begin{eqnarray}}
\newcommand{\eea}{\end{eqnarray}}

\begin{document}
\def\thefootnote{\fnsymbol{footnote}}


\begin{center}
{\large {\bf The Price of Shifting the Hubble Constant}}

\bigskip

\bigskip

{\large \noindent  Jarah
Evslin${}^{1,2}$\footnote{jarah@impcas.ac.cn}, Anjan A. Sen${}^{3}$\footnote{aasen@jmi.ac.in} and Ruchika${}^{3}$\footnote{ruchika@ctp-jamia.res.in} }



\vskip.7cm

\vspace{0em} {\em  1) Institute of Modern Physics, CAS, NanChangLu 509, Lanzhou 730000, China\\
2) University of the Chinese Academy of Sciences, YuQuanLu 19A, Beijing 100049, China\\
3) Centre for Theoretical Physics, Jamia Millia Islamia, New Delhi-110025, India}

\vskip .4cm

\end{center}

\noindent
\begin{center} {\bf Abstract} \end{center}

\noindent
An anisotropic measurement of the baryon acoustic oscillation (BAO) feature fixes the product of the Hubble constant and the acoustic scale $H_0 r_d$.  Therefore,  regardless of the dark energy dynamics, to accommodate a higher value of $H_0$ one needs a lower $r_d$ and so necessarily a modification of early time cosmology.  One must either reduce the age of the Universe at the drag epoch or else the speed of sound in the primordial plasma.  The first can be achieved, for example, with dark radiation or very early dark energy, automatically preserving the angular size of the acoustic scale in the Cosmic Microwave Background (CMB) with no modifications to post-recombination dark energy.  However it is known that the simplest such modifications fall afoul of CMB constraints at higher multipoles.  As an example, we combine anisotropic BAO with geometric measurements from strong lensing time delays from H0LiCOW and megamasers from the Megamaser Cosmology Project to measure $r_d$, with and without the local distance ladder measurement of $H_0$.  We find that the best fit value of $r_d$ is indeed quite insensitive to the dark energy model, and is also hardly affected by the inclusion of the local distance ladder data.



\pagebreak

\renewcommand{\thefootnote}{\arabic{footnote}}
\setcounter{footnote}{0}

\section{Introduction}
\subsection{Motivation}

The $\Lambda$CDM cosmological model has become the Standard Model of cosmology in the past 20 years in part because its remarkably simple treatment of cosmological acceleration \cite{perlmutternatura,riess98,perlmutter}, as the result of a cosmological constant, has been confirmed time and time again with ever greater precision and at various redshifts by multiple cosmological probes.  
On the other hand, through its consistency with various cosmological properties, while remaining in a regime which is generally believed to allow a semiclassical treatment (but see Ref.~\cite{brandenbergerinflazione}), the inflationary paradigm has won the hearts of most cosmologists and so has been effectively incorporated into this Standard Cosmological Model.  Inflation also comes with cosmological acceleration, but the origin is the condensate of a new field, the inflaton, and not a cosmological constant.  Although the inflaton has never been observed, many other such condensates are known to exist in the electroweak and strongly interacting sectors of the Standard Model of particle physics.   The fact that known condensates do not yield acceleration, as would be expected by minimally coupling gravity to the Standard Model, is usually attributed to an unknown mechanism which somehow does not prevent the inflaton from causing acceleration.  

Despite these theoretical gaps, a consistent picture has emerged.  The Universe is accelerating now and this acceleration is well described by a cosmological constant, which fits the expansion data extraordinarily well at $z<1$, and it also underwent many efolds of acceleration at very high redshift.   Given these bouts of expansion at early and late times, one may wonder about acceleration in the middle \cite{prestissimo}.  Clearly efolds of intermediate time acceleration are observationally excluded, but in the age of precision cosmology even a few percent of anomalous expansion,  which could be called early dark energy whether before or after recombination, could skew parameter measurements beyond their reported uncertainties and lead to tension between various datasets.   Ref.~\cite{kampresto} has shown that one such model can lead to a shift of the best fit $H_0$ of 1.6 km/sec/Mpc with no other deviations from $\Lambda$CDM invoked, while some older models satisfy looser bounds \cite{amendolapresto}.

Early dark energy is just one of many examples of new physics which would affect the cosmological expansion rate.  While some parametrizations of interactions between dark energy and dark \cite{wetterich,anjan2002} or even visible matter \cite{amendoladarkcoppia} or dark energy itself \cite{viscous} have been studied, in general dark matter may also simply decay to radiation with a lifetime of order $10^{11}$ years, which would cause a deficit in the dark matter today as compared with that at, say $z>1$ as well as nonneglible dark radiation today.  On the other hand, in solitonic dark matter models, dark matter is nucleated, for example via the Kibble mechanism, and so dark energy (gravitating or not) is turned into dark matter before recombination \cite{monopoli,monopoli2}.  Similarly in Bose Einstein Condensate dark matter models \cite{sinbec,leekohbec}, the dark matter condenses at some early time.  These models predict a dark matter surplus at recent times as compared with that before recombination, as does the interacting dark matter-dark energy model of Ref.~\cite{valentinointer} when fit to recent cosmological datasets.
   
The cosmological Standard Model, like the Standard Model of particle physics and that of neutrino physics, has its 2$\sigma$ and 3$\sigma$ anomalies.  History shows that most of these anomalies disappear \cite{pisky}, nonetheless some may lead the way beyond $\Lambda$CDM and so their study is a central topic in each field.  The upshot of the above meandering discussion is that it is hopeless to parameterize all of the possible deviations from $\Lambda$CDM and conversely, given a short list of favored deviations from $\Lambda$CDM it is unlikely that the resolution to a given anomaly will lie on the list\footnote{That said, the sum of the neutrino masses is unknown and is likely a constant of Nature (but see Ref.~\cite{dvalineut,hanndyn}) and so it is an inevitable parameter in any model.}.

\subsection{Methodology}

This motivates studies of these anomalies which are as model-independent as possible.  Therefore we will restrict our attention to geometric observables, which are less prone to some systematics arising from stellar evolution and dust extinction, although in general they are affected by modelling assumptions.  In particular, following the logic of Ref.~\cite{heavens} we will assume only homogeneity, isotropy, minimal coupling of electromagnetism to gravity (so that light follows geodesics) and that the BAO length scale $r_d$ is position independent in comoving coordinates to determine $r_d H(z)$ and weighted combinations thereof from anisotropic BAO measurements.  

Were $H(z)$  known directly, one could immediately determine $r_d$.  This is not the case.  Instead we will use the weighted combination of $H(z)$ values which appears in the strong lensing time delay distances determined by the H0LiCOW collaboration.  These will be combined with the angular diameter distances to masers from the Megamaser Cosmology Project, which largely constrain $H_0$.  To bind these disparate measurements of combinations of $H(z)$'s together, we will assume that the evolution of the dark energy density is sufficiently smooth to be well approximated at these low redshifts by various simple parameterizations such as the linear CPL form \cite{cpl1,cpl2}.  

This assumption may appear to be in contradiction with the preceding diatribe on the needed generality in a study of dark energy, but actually this assumption is only imposed at the redshifts of the data considered.  In particular, at $z>1$ only one set of anisotropic BAO data is used, and it has a small effect on the best fits, and so this assumption essentially only constrains the dark energy behavior at $z<1$ where any gross violation would likely be apparent in 1a supernova Hubble diagrams\footnote{Indeed, Refs.~\cite{riesstrouble,kaplinghat17} performed a similar calculation using a more general form for the dark energy dynamics and found that, although in some cases the BAO data implies a very sharp dark energy evolution at very low $z$ (found also in Ref.~\cite{sharp}), once supernova data is included this preference disappears.  In general this evolution occurs at lower redshifts than the strongest BAO constraints, so that it compensates for a mismatch between the Hubble parameter at BAO measurement redshifts, using an externally determined $r_d$, and direct $H_0$ measurements.}.  Therefore our results would be minimally affected by any change in the dark energy dynamics at $z>1$ and would be entirely unaffected by any dark energy dynamics at $z>2.4$.

We will then use a straightforward frequentist likelihood analysis to combine these measurements into profile likelihoods for $r_d$, with nuisance parameters chosen to maximize the likelihood function.  The resulting value will be between 135 and about 140 Mpc depending in the data set and dark energy model.  This is somewhat lower than the best fit Planck results \cite{planck} in the case of a $\Lambda$CDM cosmology, which we adopt as a benchmark.   We find that our best fit value is somewhat robust against changes in the dark energy model, and even changes in the dataset so long as BAO is included.  However we find that the inclusion of the local distance ladder measurement of $H_0$ from Ref.~\cite{riess}, as has been done in Ref.~\cite{riesstrouble}, reduces the uncertainty in $r_d$ and so increases the statistical significance of this incompatibility.  

 The evidence for a model which is qualitatively similar to $\Lambda$CDM is quite overwhelming, and so we will argue that the redshift to the drag epoch at which $r_d$ measures the acoustic length scale is robust at the level of precision of these likelihoods.   Therefore the low $z$ measurements determine the metric size of the acoustic feature at the drag epoch and find it to be smaller than the benchmark Planck value.  
 
These Planck results assume a $\Lambda$CDM model and we do not investigate how a change in the dark energy model would affect the Planck best fit, a task which we leave to the sequel.   This is because our goal in the present paper is not to test the analysis of Planck data nor to fit the CMB data to our models, but rather to use probes of late time cosmology to learn about $r_d$. CMB measurements, being sensitive directly to a number of features of early time cosmology, are not suitable for this task.  The role of the the Planck $\Lambda$CDM best fit in this paper is simply as a benchmark:   As the Planck best fit value is widely used for a number of cosmological and astrophysical applications, we feel that it is useful to test it against various cosmological datasets and models even if the models are not those from which this Planck best fit is derived.   Perhaps more to the point, there are many ways in which a smaller $r_d$ could be achieved by modifying early Universe cosmology.  These are equivalent from the point of view of our late time geometric probes, however they affect the CMB differently.  Therefore a study of their effects on the CMB would need to consider many rather different cases, and even so is unlikely to be exhaustive.
 
If the best fit value of $r_d$ derived here from BAO, lensing and megamaser data is to be realized in Nature, then one requires a change in early time cosmology.  In particular the change in early time cosmology must either reduce the age of the Universe at recombination or else it must reduce the speed of sound of the primordial plasma.  This explains the observation in Refs.~\cite{12silk,12linder} that, while modifications in the dark energy equation of state may reconcile tension between CMB and local distance ladder measurements, no such change to late time cosmology removes the tension when BAO is included, a result that has been extended to late time dark matter-dark energy interactions in Ref.~\cite{xiainter}.  This situation is in contrast with the tension between the BAO Lyman $\alpha$ forest and galaxy clustering datasets, which lies entirely in the late time expansion of the Universe \cite{bao2} and need not affect, for example, the angular diameter distance to recombination.

This of course will beg the question of just what modifications would be consistent with the CMB data.  The line of argument of the current paper is deductive and quite model independent: from certain assumptions and given certain data we establish necessary and sufficient conditions for consistency with BAO, strong lensing and megamaser data.  However, given the wealth of modifications to early time cosmology that satisfy this criteria, we know of no such model independent conclusions regarding the CMB and so will only make some general remarks considering specific CMB constraints and will leave model-building to the sequel.

\section{Data} \label{datisez}

\subsection{BAO Data}

We combine isotropic BAO measurements from the 6dF Survey \cite{6df}, at an effective redshift of $z=0.106$, with the reconstructed Sloan Digital Sky Survey Data Release 7 \cite{dr7} main galaxy sample (MGS) at an effective redshift of $z=0.15$ and the extended Baryon Oscillation Spectroscopic Survey (eBOSS) quasar clustering at $z=1.52$ \cite{ebossquasar} with the anisotropic measurements from the BAO only analysis of the Baryon Oscillation Spectroscopic Survey (BOSS) analysis \cite{dr12} and Lyman $\alpha$ forest samples \cite{lyadr12} at effective redshifts of $z=0.38$, $z=0.51$, $z=0.61$ and $z=2.4$. 

\begin{table}
\centering
\begin{tabular}{|c|l|l|l|l|}
\hline
Data Set&Redshift&Constraint&Ref.\\
\hline\hline
6dF&$z=0.106$&$\frac{D_V(0.106)}{r_d}=2.98\pm 0.13$&\cite{6df}\\
\hline
MGS&$z=0.15$&$\frac{D_V(0.15)}{r_d}=4.47\pm 0.17$&\cite{dr7}\\
\hline
eBOSS quasars&$z=1.52$&$\frac{D_V(1.52)}{r_d}=26.1\pm 1.1$&\cite{ebossquasar}\\
\hline
\end{tabular}
\caption{Isotropic BAO scale measurements used in this analysis.}
\label{isobaotab}
\end{table}

\begin{table}
\centering
\begin{tabular}{|c|l|l|l|l|}
\hline
Data Set&Redshift&Constraint&Ref.\\
\hline\hline
BOSS DR12&$z=0.38$&$\frac{D_A(0.38)}{r_d}=7.42$&\cite{dr12}\\
\hline
BOSS DR12&$z=0.38$&$\frac{D_H(0.38)}{r_d}=24.97$&\cite{dr12}\\
\hline
BOSS DR12&$z=0.51$&$\frac{D_A(0.51)}{r_d}=8.85$&\cite{dr12}\\
\hline
BOSS DR12&$z=0.51$&$\frac{D_H(0.51)}{r_d}=22.31$&\cite{dr12}\\
\hline
BOSS DR12&$z=0.61$&$\frac{D_A(0.61)}{r_d}=9.69$&\cite{dr12}\\
\hline
BOSS DR12&$z=0.61$&$\frac{D_H(0.61)}{r_d}=20.49$&\cite{dr12}\\
\hline
BOSS DR12&$z=2.4$&$\frac{D_A(2.4)}{r_d}=10.76$&\cite{lyadr12}\\
\hline
BOSS DR12&$z=2.4$&$\frac{D_H(2.4)}{r_d}=8.94$&\cite{lyadr12}\\
\hline
\end{tabular}
\caption{Anisotropic BAO scale measurements used in this analysis.}
\label{anisobaotab}
\end{table}

The isotropic and anisotropic BAO measurements are summarized in Tables~\ref{isobaotab} and \ref{anisobaotab} respectively.  The BAO only analysis of Ref.~\cite{dr12} is used, which only uses the peak location and not the matter power spectrum shape.  The uncertainties for this analysis were not reported in that reference but are available from the SDSS website in the file {\it{BAO\_consensus\_covtot\_dM\_Hz.txt}}.  Combined with the covariance matrix for the Lyman $\alpha$ BAO detection in Ref.~\cite{lyadr12}, the covariance matrix corresponding to the data in Table~\ref{anisobaotab} is
\beq
{\mathbf{C}}=\left(
\begin{tabular}{cccccccc}
0.0150&-0.0358&0.0071&-0.0100&0.0032&-0.0036&0&0\\
-0.0357&0.5304&-0.0160&0.1766&-0.0083&0.0616&0&0\\
0.0071&-0.0160&0.0182&-0.0323&0.0097&-0.0131&0&0\\
-0.0100&0.1766&-0.0323&0.3267&-0.0167&0.1450&0&0\\
0.0032&-0.0083&0.0097&-0.0167&0.0243&-0.0352&0&0\\
-0.0036&0.0616&-0.0131&0.1450&-0.0352&0.2684&0&0\\
0&0&0&0&0&0&0.1358&-0.0296\\
0&0&0&0&0&0&-0.0296&0.0492\\
\end{tabular}
\right).
\eeq 
This data is combined into a $\chi^2$ statistic
\beq
\chi^2_{\rm{BAO}}=\chi^2_{\rm{iso}}+\chi^2_{\rm{aniso}}\hsp
\chi^2_{\rm{iso}}=\sum_i\left(\frac{v_i-d^{\rm{iso}}_i}{\sigma_i}\right)^2\hsp
\chi^2_{\rm{aniso}}=\left(\mathbf{w-d}^{\rm{aniso}}\right)^\perp \mathbf{C}^{-1} \left(\mathbf{w-d}^{\rm{aniso}}\right) \label{chibao}
\eeq
where the vectors $\mathbf{d}^{\rm{iso}}$ and $\mathbf{d}^{\rm{aniso}}$ are the isotropic and anisotropic best fit data from Tables~\ref{isobaotab} and \ref{anisobaotab} and $\mathbf{v}$ and $\mathbf{w}$ are the predictions for these vectors in a given cosmological model.  The $\sigma_i$ are the uncertainties from Table~\ref{isobaotab}.  Due to the relatively small number of mocks, the Lyman $\alpha$ forest BAO likelihood profile deviates noticeably from the Gaussian form used here when a model is more than about 2$\sigma$ from the data.  However this dataset will contribute to our fits essentially only by shifting the best fit parameters for the dark energy evolution, and once this shift has occurred the model will lie within 2$\sigma$ of the data, and so this non-Gaussianity is inconsequential for our study of $H_0$.  It would be relevant if we were searching for evidence for dark energy dynamics or if we used the likelihoods of the dark energy parameters far from their maxima.

\subsection{Other Data}

The BAO data does not directly constrain $r_d$, $H_0$ or even $H(z)$, rather it constrains various weighted averages of $H(z) r_d$.  Our goal is to use low redshift observations to measure $r_d$, therefore we need independent constraints on $H(z)$.  There are a number of cosmological probes which are sensitive to $H(z)$ (for example the cosmic chronometers of Ref.~\cite{chron}).  We will largely restrict our attention to geometric probes, which determine $H(z)$ using geometry, as these are independent of many of the astrophysical and cosmological assumptions that may affect the others.  For example, they are not affected by stellar evolution, dust extinction or variations in local metalliticity.  Most importantly for our analysis, they are reasonably robust, as compared with the present uncertainties, against changes in the cosmological model at higher redhshifts than the observations themselves.

There remain of course some sources of systematic error which are difficult to quantify, in particular regarding modelling.  For example, in the case of the Lyman $\alpha$ BAO analysis, one needs to estimate the continuum component of the Lyman $\alpha$ absorption, so that it may be subtracted to obtain the forest and extract the corresponding mass density.    In the near future eBOSS will complete its anisotropic BAO analysis of the quasar-quasar correlation function at similar redshfits \cite{gongboforecast} allowing an independent test of these BAO measurements.

In the case of strong lensing time delays, the determination of the time delay distance requires a determination of the density profile of the lens and the line of sight \cite{wilsonlens}, but there are several potentially dangerous degeneracies in the construction of the lens model \cite{falco85,ss} from the observed source positions.  To break these degeneracies in the determination of the lens profile, one generally uses the line of sight velocity dispersion.  However, given the line of sight velocity dispersion there is an exact degeneracy between the density profile and the unknown velocity anisotropy \cite{degen}.  In principle, without a measurement of the velocity anisotropy, which is infeasible at cosmological distances, the density profile cannot be constrained and thus the cosmological parameter estimates may be biased \cite{ssmass,kamionkowski}.  The H0LiCOW collaboration treats this problem in Ref.~\cite{holy4} by assuming that the velocity anisotropy has a certain functional dependence on the radius, with one free parameter, and then that parameter is marginalized using a uniform prior with a fixed range. Needless to say, both the choice of parameterization and the prior on the parameter are assumptions, leading to contributions to the error budget which are difficult to quantify.   In the future, some improvement will come from spatially resolved kinematics \cite{tommaso17}.  

This problem will be ameliorated in the era of extremely large telescopes, which will provide both high resolution images of these lens systems \cite{tmt} and also proper motion measurements of stars in galaxies of various morphologies as far as the Virgo cluster.  One can then motivate an anisotropy profile for the lens system by comparing it to similar, nearby systems whose anisotropies have been measured.  

To determine the mass distribution along the line of sight, Ref.~\cite{holy3} uses such mass distributions in $\Lambda$CDM simulations.  This will eventually limit the applicability of high precision strong lensing time delays to cosmologies which are sufficiently similar to the fiducial cosmologies of the simulations.

In the case of water megamasers, modelling the maser itself is essential.  For example. if the number of rings is determined incorrectly, or if part of the signal is interpreted as arising from the wrong ring, this would lead to an error which is unlikely to be quantified by the assigned error budget.  This is a problem that in principle can be solved with deeper, higher resolution images.

\begin{table}
\centering
\begin{tabular}{|c|l|l|l|l|l|l|}
\hline
Lens&$z_d$&$z_s$&$\mu_d$&$\sigma_D$&$\lambda_D$&Ref.\\
\hline\hline
B1608+656&$0.6304$&$1.394$&$7.0531$&$0.22824$&$4000.0$&\cite{suyu2010}\\
\hline
RXJ1131-1231&$0.295$&$0.654$&$6.4682$&$0.20560$&$1388.8$&\cite{suyu2014}\\
\hline
HE0435-1223&$0.4546$&$1.693$&$7.5793$&$0.10312$&$653.9$&\cite{holy4}\\
\hline
\end{tabular}
\caption{Strong lensing time delays used in this analysis.}
\label{lenstab}
\end{table}

The main dataset which we will use is the collection of strong lensing time delays produced by the H0LiCOW collaboration.  This program is described in Ref.~\cite{holy1} and the data is summarised in Table 3 of Ref.~\cite{holy5} and is repeated in Table~\ref{lenstab} of the present note.  Here $z_s$ and $z_d$ are the redshift of the source quasar and the lensing galaxy respectively.   The parameters $\lambda_D$, $\mu_D$ and $\sigma_D$ determine the likelihood of a given time delay distance $D_{\Delta t}=x$ as
\beq
P(x)=\frac{1}{\sqrt{2\pi}\left(x-\lambda_D\right)\sigma_D}\rm{exp}\left[-\frac{\left({\rm{ln}}\left(x-\lambda_D\right)-\mu_D\right)^2}{2\sigma_D^2}
\right].
\eeq
By an abuse of notation we will simply refer to $-2$ times the log likelihood as $\chi^2_{\rm{lens}}$
\beq
\chi^2_{\rm{lens}}=-2\ {\rm{ln}}(P(x))=\frac{\left({\rm{ln}}\left(x-\lambda_D\right)-\mu_D\right)^2}{\sigma_D^2}+2\ {\rm{ln}}\left[\sqrt{2\pi}\left(x-\lambda_D\right)\sigma_D
\right] .
\eeq

We will also use angular diameter distances measured using water megamasers under the Megamaser Cosmology Project \cite{mcp1}.  We do not include the uncertainty in the distance caused by the peculiar velocity, as this has a negligible effect on our results.  The data are summarized in Table~\ref{masertab}.   The $\chi^2$ statistic $\chi^2_{\rm{maser}}$ will be calculated using the same standard formula as is used for $\chi^2_{\rm{iso}}$ in Eq.~(\ref{chibao}).  We will not use the megamaser NGC 6323 \cite{megano} or the Hubble parameter determination from the standard siren GW170817 \cite{gw2017,gw2017b} because, due to their large uncertainties, they would have a negligible effect on our results.   We note that these standard siren measurements are not purely geometric in that it is assumed that the source galaxy has been correctly identified and, more importantly, a 10\% correction to the recessional velocity is applied to correct for the peculiar velocity caused by the local gravitational field.  In addition the latter reference uses an estimate of the inclination angle based on a model of the system's jet.

\begin{table}
\centering
\begin{tabular}{|c|l|l|l|l|}
\hline
Maser&Redshift&Constraint&Ref.\\
\hline\hline
UGC 3789&$z=0.0116$&$\frac{D_A(0.0116)}{\rm{Mpc}}=49.6\pm 5.1$&\cite{mcp4}\\
\hline
NGC 6264&$z=0.0340$&$\frac{D_A(0.0340)}{\rm{Mpc}}=144\pm 19$&\cite{mcp5}\\
\hline
NGC 5765b&$z=0.0277$&$\frac{D_A(0.0277)}{\rm{Mpc}}=126.3\pm 11.6$&\cite{mcp8}\\
\hline
\end{tabular}
\caption{Megamaser measurements used in this analysis.}
\label{masertab}
\end{table}

Finally, we will sometimes include the distance ladder measurement $H_0=73.24\pm 1.74$ km/sec/Mpc from Ref.~\cite{riess}.  It is not an entirely geometric measurement, even if it uses megamasers in part, however it relies upon astrophysical instead of cosmological assumptions and is independent of early time cosmology.

\section{Results}
\subsection{Models}
\begin{table}
\centering
\begin{tabular}{|c|l|l|}
\hline
Cosmology&Parameters&Constraints\\
\hline\hline
$\Lambda$CDM-Planck&$r_d$&$\Omega_m=0.31,\ P=30.0,\ w(z)=-1$\\
\hline
H$\Lambda$CDM-Planck&$r_d,\ H_0$&$\Omega_m=0.31,  w(z)=-1$\\
\hline
$\Lambda$CDM&$r_d,\ \Omega_m,\ H_0$&$w(z)=-1$\\
\hline
$w$CDM&$r_d,\ \Omega_m,\ H_0,\ w$&$w(z)=w$\\
\hline
CPL&$r_d,\ \Omega_m,\ H_0,\ w_0,\ w_a$&$w(z)=w_0+\frac{z}{1+z}w_a$\\
\hline
\end{tabular}
\caption{Five cosmological models}
\label{modtab}
\end{table}

We consider the five cosmological models summarized in Table~\ref{modtab}.  The first is the standard, flat $\Lambda$CDM cosmology with the parameters $\Omega_m$ and
\beq
P=\frac{c}{H_0 r_d}
\eeq
fixed to their best fit values from the Planck experiment \cite{planck} including lensing and polarization data.  Note however that we do not fix $r_d$, since our goal is to determine a likelihood for $r_d$.  In the second, $H_0$ is freed.  The third is also a flat $\Lambda$CDM model with dark energy equation of state $w(z)=-1$, but $r_d$, $P$ and $\Omega_m$ unconstrained.  Next, we consider the same model but allow the dark energy equation of state $w(z)$ to assume any $z$-independent value $w$.  Finally, we consider a flat cosmology with $w(z)$ of the CPL form \cite{cpl1,cpl2}
\beq
w(z)=w_0+\frac{z}{1+z}w_a.
\eeq

\subsection{Analysis}
In a spatially flat Universe with electromagnetism minimally coupled to gravity, the various cosmological distances described above can be written
\bea
D_A(z)&=&\frac{c}{1+z}\int_0^z\frac{dz^\prime}{H(z^\prime)} \label{obs}\\
D_H(z)&=&\frac{c}{H(z)}\nonumber\\
D_V(z)&=&\left( (1+z)D_A(z)\right)^{2/3}\left(zD_H(z)\right)^{1/3}\nonumber\\
D_{\Delta t}(z_d,z_s)&=&c\frac{\int_0^{z_d}\frac{dz^\prime}{H(z^\prime)}\int_0^{z_s}\frac{dz^\prime}{H(z^\prime)}}{\int_{z_d}^{z_s}\frac{dz^\prime}{H(z^\prime)}}\nonumber
\eea
where $c$ is the speed of light and the function $H(z)$ is given by
\beq
H(z)=H_0 \sqrt{\Omega_m(1+z)^3+(1-\Omega_m)e^{3\int_0^z\frac{1+w(z^\prime)}{1+z^\prime}dz^\prime}}
\eeq
in terms of the Hubble parameter $H_0$, the matter fraction of the critical density $\Omega_m$ and the equation of state of dark energy $w(z)$.  Using these formulas, for a given cosmological model in Table~\ref{modtab} one can find the observables (\ref{obs}) and so can calculate the $\chi^2$ functions and log likelihoods described in Sec.~\ref{datisez}.

The above formula for $D_{\Delta t}(z_d,z_s)$ was derived from the usual formula in terms of angular diameter distances
\beq
D_{\Delta t}(z_d,z_s)=(1+z_d)\frac{D_dD_s}{D_{ds}}
\eeq
using the identities 
\beq
D_s=\frac{c}{1+z_s}\int_0^{z_s}\frac{dz^\prime}{H(z^\prime)}\hsp
D_d=\frac{c}{1+z_d}\int_0^{z_d}\frac{dz^\prime}{H(z^\prime)}\hsp
D_{ds}=\frac{c}{1+z_s}\int_{z_d}^{z_s}\frac{dz^\prime}{H(z^\prime)}.
\eeq

We are interested in the value of $r_d$ consistent with various datasets, since a modification of $r_d$ requires a modification of early time cosmology and we want to know whether early time cosmology indeed needs to be modified to accommodate current data, in contrast with claims in the literature \cite{linderdemodel}.  However the models at hand contain a number of other parameters.  In line with our desired level of model-independence, we do not wish to impose priors on these nuisance parameters.  Therefore we will opt for a frequentist analysis.

We will use the profile likelihood, which has many of the same properties as the likelihood function itself~\cite{patefield}, even if one allows the full set of functions $w(z)$ instead of parametrizing it \cite{murphy} as is done here.  The profile likelihood function is obtained by simply maximising the likelihoods of the nuisance parameters, or equivalently minimizing the values of $\chi^2$.   More precisely, for each cosmological model, and each set of datasets, we will add the corresponding log likelihoods.  Then for each value of $r_d$ we will choose all of the other parameters so as to maximize this sum.  The result is the profile log likelihood for $r_d$.  We will double this so that we get an estimator which is roughly speaking a $\chi^2$ statistic.  We will report the quantity $\Delta\chi^2$, which is minus twice the profile log likelihood minus its own minimal value.    We report this quantity because Wilks' theorem guarantees that $\Delta\chi^2$ will nearly follow a $\chi^2$ distribution with one degree of freedom.

We also estimate the 1$\sigma$ uncertainty on $r_d$ as half of the size of the interval at which this  $\Delta\chi^2$ is less than one.   The corresponding interval is the range of values of $r_d$ for which there exists a set of nuisance parameters that yields a $p$ value greater than 0.32.  Note that the nuisance parameters are optimized, not marginalized, in this analysis and so no priors are needed.

\subsection{Results}

\noindent
{\bf{BAO only}}

Using only the BAO data, one can determine $c/(H(z) r_d)$ but it is not possible to separate $H(z)$ and $r_d$.  Therefore there is no difference between the compatibility of the data with the $\Lambda$CDM-Planck and H$\Lambda$CDM-Planck models, as $r_d$ and $H_0$ are separately unconstrained.  Given a dark energy model at the redshifts of the BAO data, one can choose the nuisance parameters to maximize the likelihood and so obtain the profile likelihood for $P=c/(H(z)r_d)$ in that model.   The simplest case is that of $\Lambda$CDM, where the only parameters which affect the likelihood are $P$ and $\Omega_m$.   

\begin{figure} 
\begin{center}
\includegraphics[width=3.7in,height=3.7in]{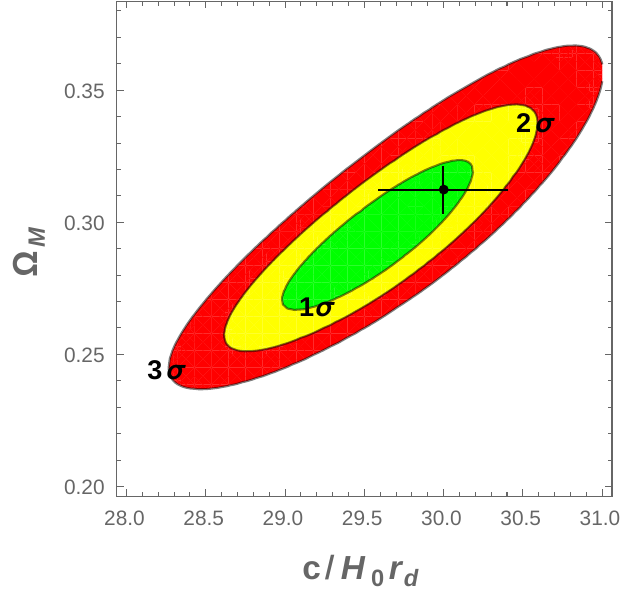}
\includegraphics[width=4.0in,height=2.8in]{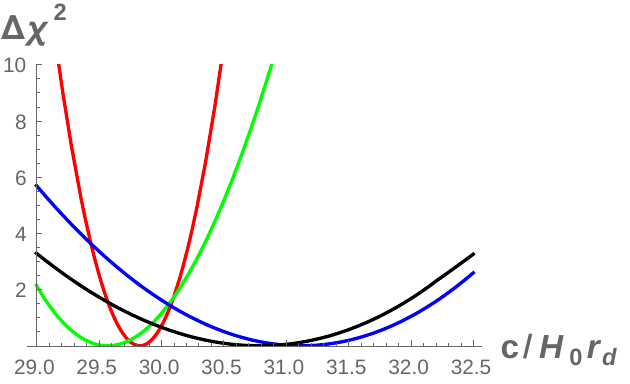}
\caption{BAO only results.  Top: In the $\Lambda$CDM model, the only relevant parameters are $\Omega_m$ and $c/(H_0r_d)$.  The standard error ellipses are plotted, corresponding to $\Delta\chi^2_{\rm{BAO}}=2.3,\ 6.0$ and $11.8$. The Planck benchmark value, with its corresponding error bars derived in the case of a $\Lambda$CDM model, is included for comparison.  Bottom: $\Delta\chi^2$ is plotted for various values of $c/(H_0 r_d)$ with all nuisance parameters optimized.  The red, green, blue and black curves correspond to the $H\Lambda$CDM-Planck, $\Lambda$CDM, wCDM and CPL models respectively.}
\label{baofig}
\end{center}
\end{figure}

The full two-dimensional $\Delta\chi^2$ on this parameter space is depicted in the first panel of Fig.~\ref{baofig}. The Planck $\Lambda$CDM best fit value \cite{planck} with its associated uncertainties is shown and excellent agreement can be seen.  The profile likelihoods for $P$ are drawn in the second panel of Fig.~\ref{baofig} and the corresponding 1$\sigma$ uncertainties are reported in Table~\ref{liktab}.

\noindent
{\bf{BAO and strong lensing}}

\begin{figure} 
\begin{center}
\includegraphics[width=3.7in,height=3.7in]{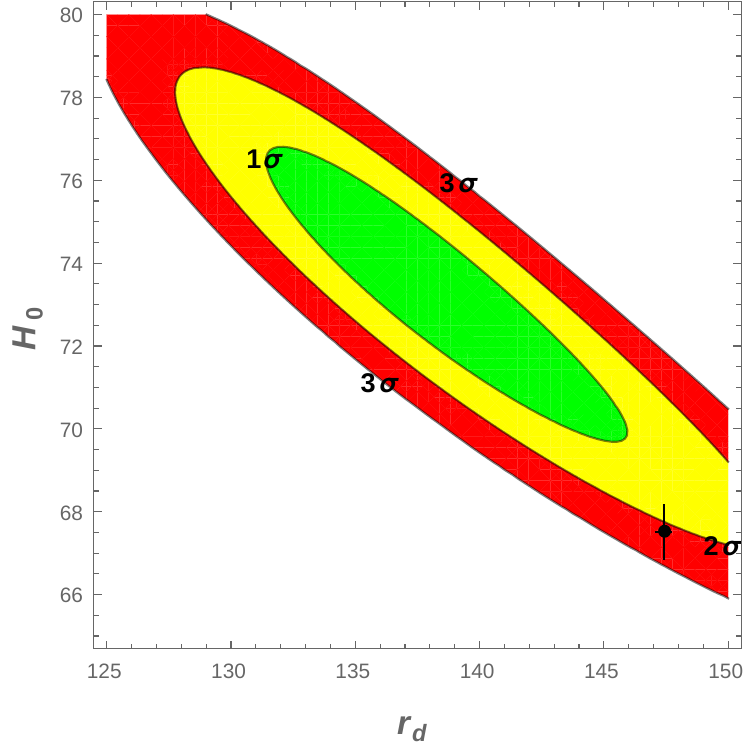}
\includegraphics[width=4.0in,height=2.8in]{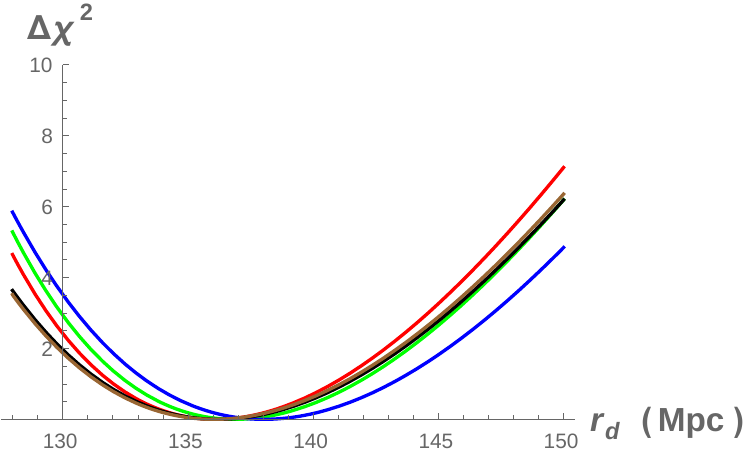}
\caption{BAO and strong lensing time delay results.  Top: In the $\Lambda$CDM model, the only relevant parameters are $\Omega_m$, $r_d$ and $H_0$.  The standard error ellipses are plotted with $\Omega_m$ optimized,  $H_0$ in units of km/s/Mpc and $r_d$ in units of Mpc.  Bottom: $\Delta\chi^2$ is plotted for various values of $H_0$ with all nuisance parameters optimized.  The red, green, blue, black and brown curves correspond to the $\Lambda$CDM-Planck, H$\Lambda$CDM-Planck, $\Lambda$CDM, wCDM and CPL models respectively. }
\label{baolfig}
\end{center}
\end{figure}

Next strong lensing time delays from H0LiCOW are included in the analysis.  Each of the three strong lenses gives a time delay distance which constrains a combination of the $H(z)$.  These are independent of $r_d$ and so $r_d$ and $H_0$ can now be separately determined.  The corresponding log likelihood, denoted $\Delta\chi^2$, of $r_d$ and $H_0$ is shown in the first panel of Fig.~\ref{baolfig} together with the Planck benchmark $H_0=67.51\pm0.64$ and $r_d=147.41\pm 0.4$.  One may observe a mild tension of about 2$\sigma$.  

The second panel shows the profile log likelihoods of the various cosmological models, where all parameters except for $r_d$ are optimized.   There is no sign that dynamical dark energy provides a better agreement with the benchmark value.  On the contrary, the only model with less than 2$\sigma$ of tension is $\Lambda$CDM.   The maximum likelihood value of $r_d$ is between 136 and 137 Mpc for all models except for $\Lambda$CDM, for which it is 138 Mpc. The corresponding 1$\sigma$ uncertainties are reported in Table~\ref{liktab}.   More general dark energy models have little effect on the uncertainty as these mostly affect the Lyman alpha BAO data, which in turn hardly influences the determination of $P$ and so $r_d$.  One sees that in each case the maximum likelihood $r_d$ is roughly 2$\sigma$ below the Planck value.

\noindent
{\bf{BAO, strong lensing and masers}}

\begin{figure} 
\begin{center}
\includegraphics[width=4.0in,height=2.8in]{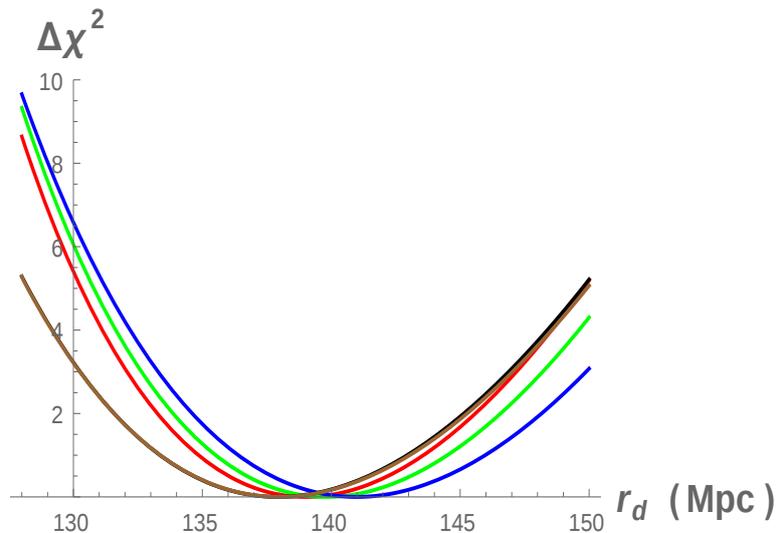}
\caption{As in the bottom panel of Fig.~\ref{baolfig}, but also including megamaser data.}
\label{baolmfig}
\end{center}
\end{figure}

Masers are included in the analysis in Fig.~\ref{baolmfig}.  As can be seen there, and in Table.~\ref{liktab}, the masers pull the best fit $r_d$  slightly towards the Planck value.  Indeed the maser data is consistent with the Planck best fit values, but the uncertainties are so large that they only lead to an increase in $r_d$ of roughly one half of a standard deviation.  Therefore at this point the mild tension lies entirely between the BAO value of $r_d H_0$, the Planck value of $r_d$ and the strong lensing time delay value of $H_0$.

\begin{figure} 
\begin{center}
\includegraphics[width=3.2in,height=2.2in]{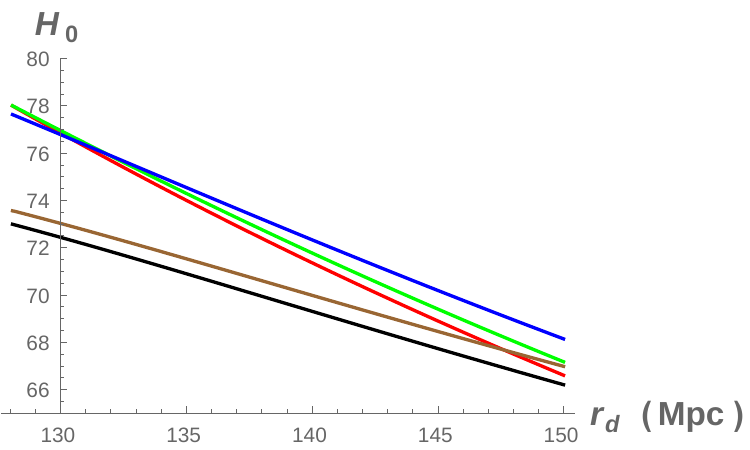}
\includegraphics[width=3.2in,height=2.2in]{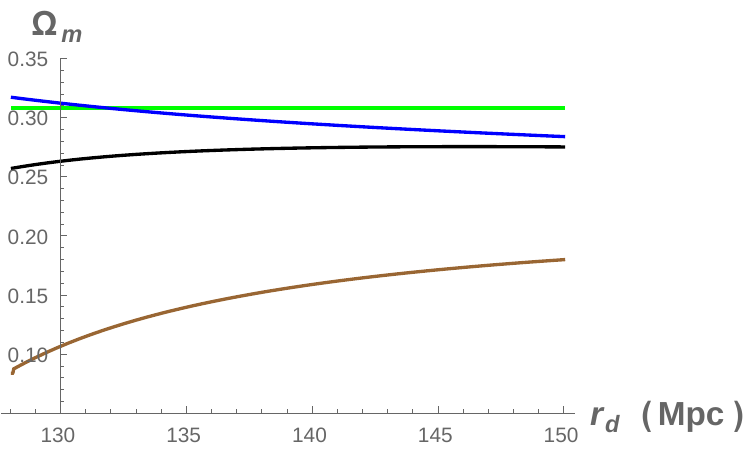}
\includegraphics[width=3.2in,height=2.2in]{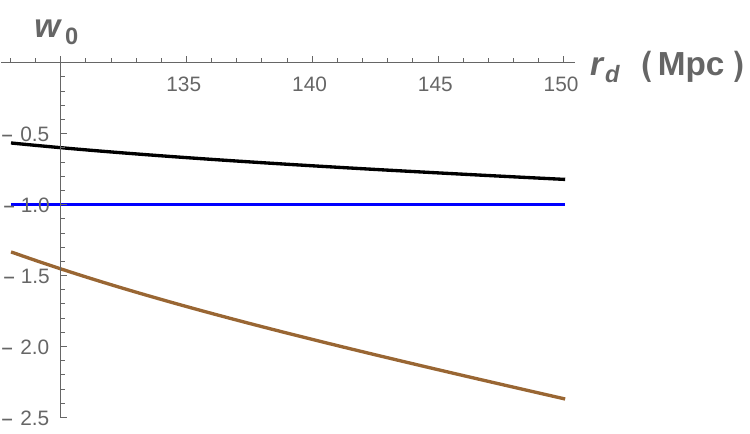}
\includegraphics[width=3.2in,height=2.2in]{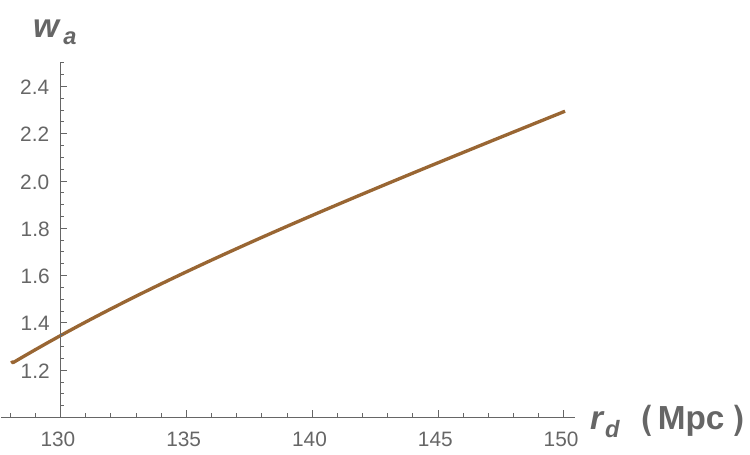}
\caption{The profiled values of the nuisance parameters in the fit of BAO, lensing and maser data.  The red, green, blue, black and brown curves correspond to the $\Lambda$CDM-Planck, H$\Lambda$CDM-Planck, $\Lambda$CDM, wCDM and CPL models respectively.}
\label{baolmpfig}
\end{center}
\end{figure}

As this analysis is our main result, we also include the nuisance parameters which maximize the likelihood at each value of $r_d$.  These are displayed in Fig.~\ref{baolmpfig}.  While the CPL best fit dark energy dynamics is quite far from a cosmological constant, one can see in Table~\ref{liktab} that it leads to essentially identical confidence intervals for $r_d$.  The role of the dynamical dark energy is largely just to satisfy the Lyman $\alpha$ BAO measurement.  This is quite different from the case of Refs.~\cite{12silk,12linder,linderdemodel} where the dynamical dark energy instead serves to change the angular diameter distance to recombination and thus to allow compatibility of the distance ladder $H_0$ with the Planck CMB $r_d$.  The strong role of BAO in our study keeps the $r_d$ far from the Planck value and so no modification to this angular diameter distance is needed.

\begin{figure} 
\begin{center}
\includegraphics[width=3.2in,height=2.2in]{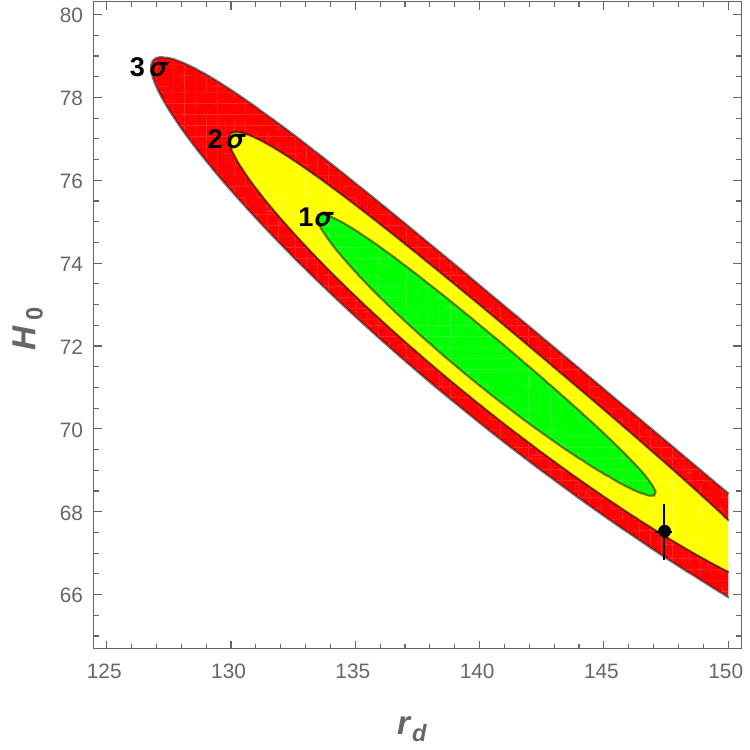}
\includegraphics[width=3.2in,height=2.2in]{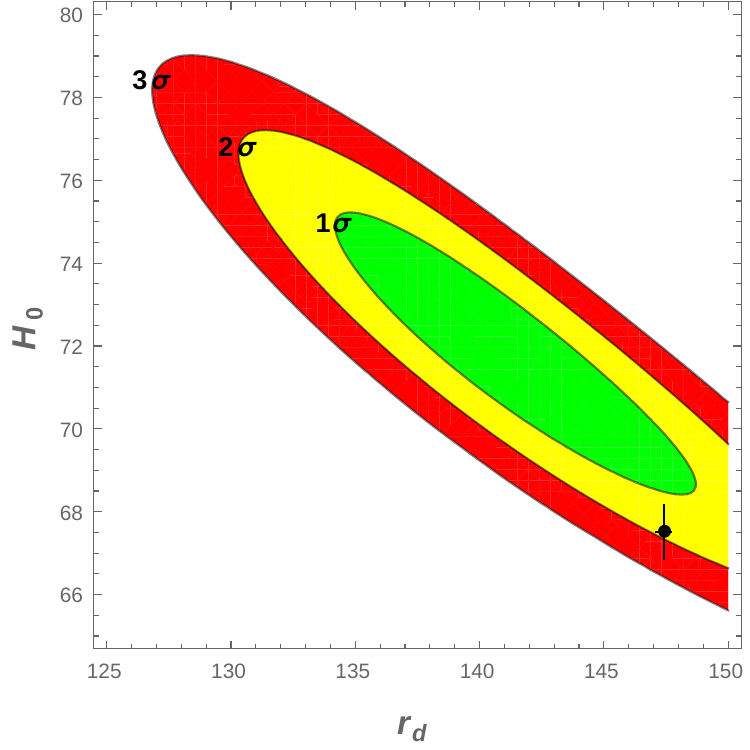}
\includegraphics[width=3.2in,height=2.2in]{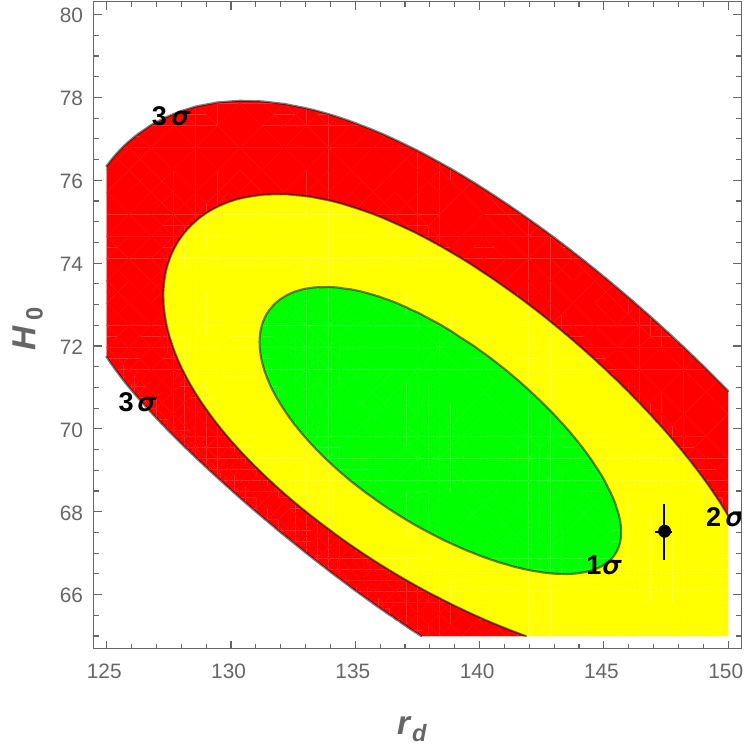}
\includegraphics[width=3.2in,height=2.2in]{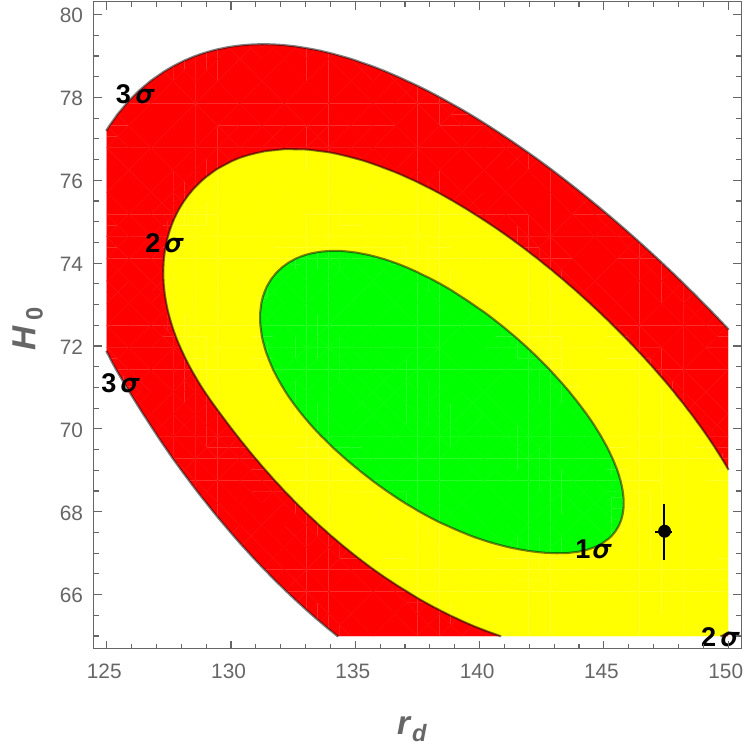}
\caption{The standard error ellipses corresponding to the profile likelihood for $H_0$ and $r_d$, with all nuisance parameters optimized.  BAO, strong lensing and megamaser data are used.  Top-left:  H$\Lambda$CDM-Planck, Top-right: $\Lambda$CDM, Bottom-left: wCDM and Bottom-right: CPL model.}
\label{baolm2dfig}
\end{center}
\end{figure}

To further illustrate this point, in Fig.~\ref{baolm2dfig} we have plotted the two-dimensional profile log likelihoods of $r_d$ and $H_0$ in each cosmological model in which $H_0$ is free.  One may observe that the level of compatibility of our best fit and the Planck $\Lambda$CDM benchmark hardly changes when dynamical dark energy is allowed.  This is despite the fact that dynamical dark energy significantly thickens the minor axis of the standard error ellipse, reducing the strong correlation between $H_0$ and $r_d$.  This is because while $r_d H_0$, the combination best constrained by BAO, is somewhat sensitive to the dark energy dynamics, the difference between the Planck $\Lambda$CDM best fit $(r_d,\ H_0)$ and the best fit here in fact lies along the least constrained direction (the semimajor axis), which appears entirely unaffected by the dark energy dynamics.  The semimajor axis constraint results from the local measurements of the Hubble constraint using strong lensing and megamasers, which are fairly insensitive to dark energy dynamics due to their low redshifts.

\noindent
{\bf{BAO, strong lensing, masers and the local distance ladder}}

\begin{figure} 
\begin{center}
\includegraphics[width=4.0in,height=2.8in]{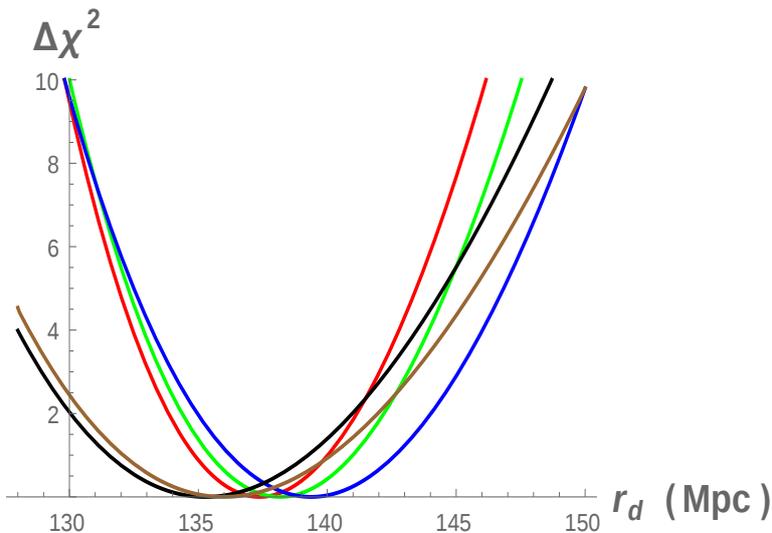}
\caption{As in the bottom panel of Fig.~\ref{baolfig}, but also including megamaser data and the local distance ladder determination of $H_0$.}
\label{baolmhfig}
\end{center}
\end{figure}

\begin{figure} 
\begin{center}
\includegraphics[width=3.2in,height=2.2in]{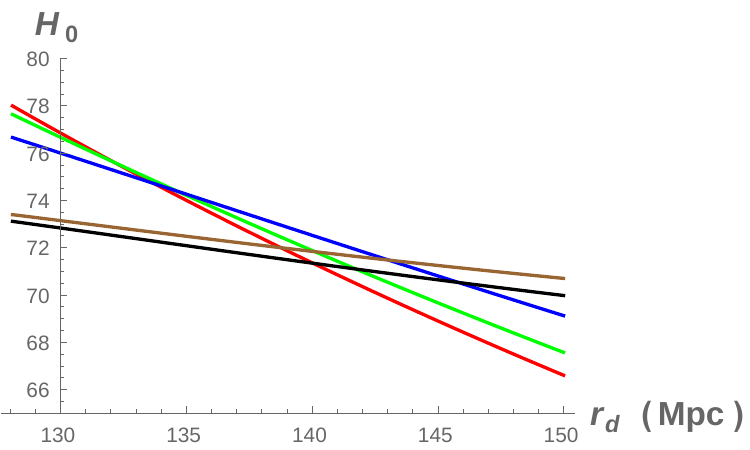}
\includegraphics[width=3.2in,height=2.2in]{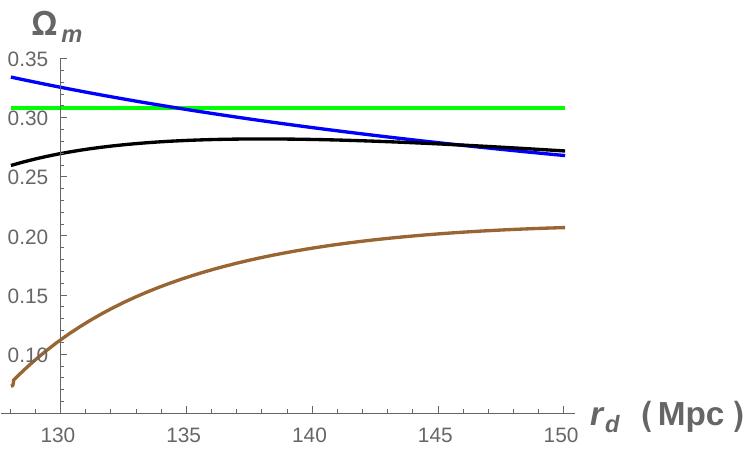}
\includegraphics[width=3.2in,height=2.2in]{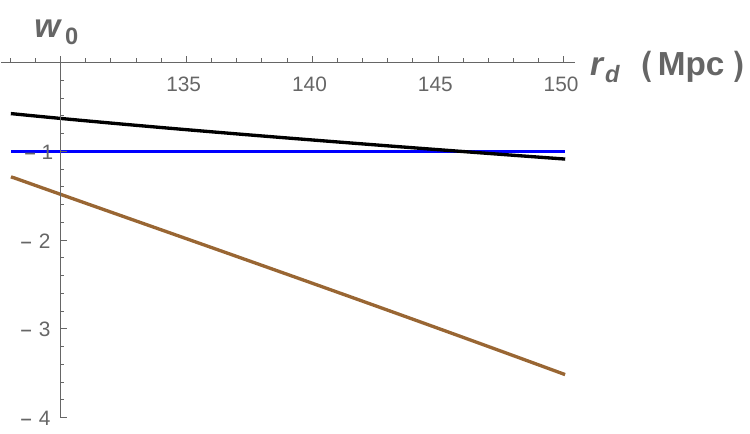}
\includegraphics[width=3.2in,height=2.2in]{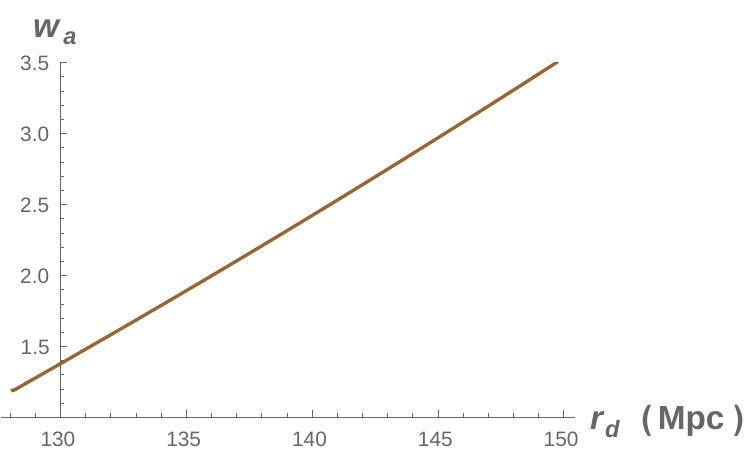}
\caption{As in Fig.~\ref{baolmpfig}, but including the local distance ladder determination of $H_0$.}
\label{baolmhpfig}
\end{center}
\end{figure}

Finally the $H_0$ from the local distance ladder is included in Figs.~\ref{baolmhfig} and \ref{baolmhpfig}.  In Table~\ref{liktab} one can see that this inclusion in fact has a only a modest effect on the best fit value of $r_d$.  This reflects the consistency between the local distance ladder and strong lensing time delay measurements of $H_0$.  The reduction in the uncertainty is between 20\% and 40\%, depending on the model.   The tension with the Planck $\Lambda$CDM benchmark values in each case remains about 3$\sigma$, increasing to 4$\sigma$ in the first two models, which incorporate some of the Planck $\Lambda$CDM parameter constraints.

\begin{figure} 
\begin{center}
\includegraphics[width=4.0in,height=2.8in]{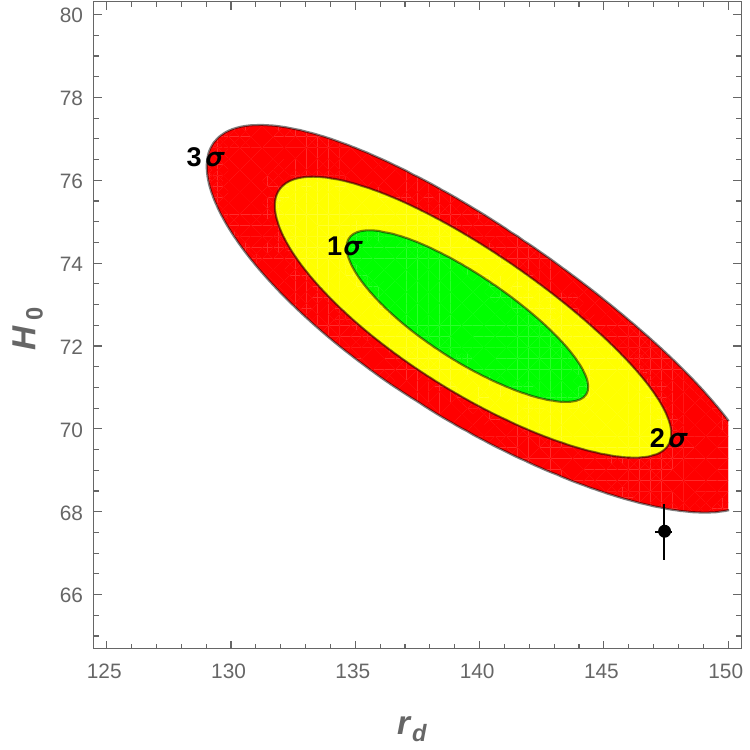}
\includegraphics[width=4.0in,height=2.8in]{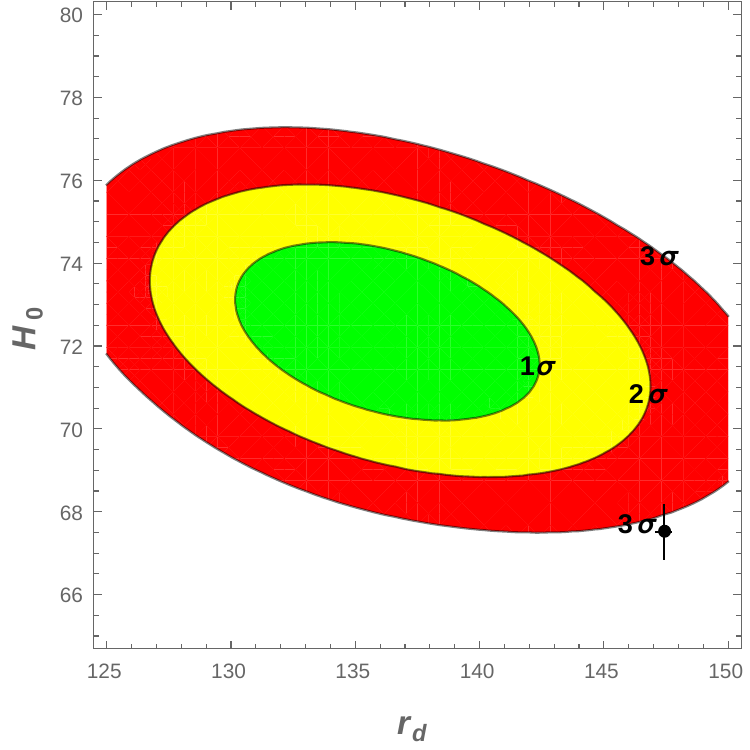}
\caption{The standard error ellipses corresponding to the profile likelihood for $H_0$ and $r_d$, with all nuisance parameters optimized.   BAO, strong lensing, megamasers and the local distance ladder are used.  Top:  $\Lambda$CDM, Bottom: CPL model.}
\label{baolmh2dfig}
\end{center}
\end{figure}

In Fig.~\ref{baolmh2dfig} we have plotted the two-dimensional profile log likelihoods of $r_d$ and $H_0$ in the $\Lambda$CDM and CPL models.   Unlike Fig.~\ref{baolm2dfig}, the tension is now greater than 3$\sigma$ in each case.  However, as in that case, one observes that dynamical dark energy does not reduce the tension.  Again dynamical dark energy thickens the minor axis of the error ellipse.  However the strong constraint on $H_0$ means that this thickening is balanced by a rotation of the ellipse towards the horizontal direction, which prevents a reduction of the tension.

\begin{table}
\centering
\begin{tabular}{|c|l|l|l|l|l|l}
\hline
Cosmology&BAO only&BAO$+$lensing&$+$masers&$+H_0$\\
\hline\hline
$\Lambda$CDM-Planck&$29.8\pm 0.2$&$136.1\pm4.7$&$139.0\pm 4.6$&$137.4\pm 2.6$\\
\hline
H$\Lambda$CDM-Planck&$29.8\pm 0.2$&$136.9\pm 4.8$&$139.8\pm 4.7$&$138.2\pm 2.8$\\
\hline
$\Lambda$CDM&$29.6\pm 0.4$&$138.0\pm 5.1$&$141.0\pm 4.6$&$139.4\pm 3.3$\\
\hline
$w$CDM&$31.1\pm 0.9$&$136.2\pm 5.1$&$137.9\pm 5.0$&$135.3\pm 4.0$\\
\hline
CPL&$30.8\pm 1.0$&$136.0\pm 5.1$&$138.0\pm 5.1$&$136.0\pm 4.2$\\
\hline
\end{tabular}
\caption{Estimates of $r_d$/Mpc with 1$\sigma$ uncertainty with various models and datasets.  In the case of BAO only, an estimate of $c/(r_d H_0)$ is reported and the first two models are equivalent.}
\label{liktab}
\end{table}

\subsection{Marginalized likelihoods}

Our analysis has used profile likelihoods, obtained by choosing the nuisance parameters so as to maximize the likelihood.  In cosmology marginalized likelihoods, in which the nuisance parameters are marginalized, are more common.  To check that our results are robust against changes in the statistical analysis, we have reproduced all of them using marginalized likelihoods computed using the Markov Chain Monte Carlo (MCMC) of Ref.~\cite{mcmc} and the code {\it{getdist}} \cite{getdist} with flat priors over intervals listed in Table~\ref{priortab}.  
\begin{table}
\centering
\begin{tabular}{|c|l|}
\hline
Quantity&Prior\\
\hline\hline
$\Omega_m$&$[0.1,0.9]$\\
\hline
$r_d$&$[130,170]$ Mpc\\
\hline
$w_0$&$[-1.9,-0.4]$\\
\hline
$w_a$&$[-4,4]$\\
\hline
$H_0$&$[50,90]$ km/sec/Mpc\\
\hline
\end{tabular}
\caption{Flat priors were used for the MCMC Bayesian analysis.  The corresponding intervals are reported in this table.}
\label{priortab}
\end{table}

\begin{figure} 
\begin{center}
\includegraphics[width=3.2in,height=2.2in]{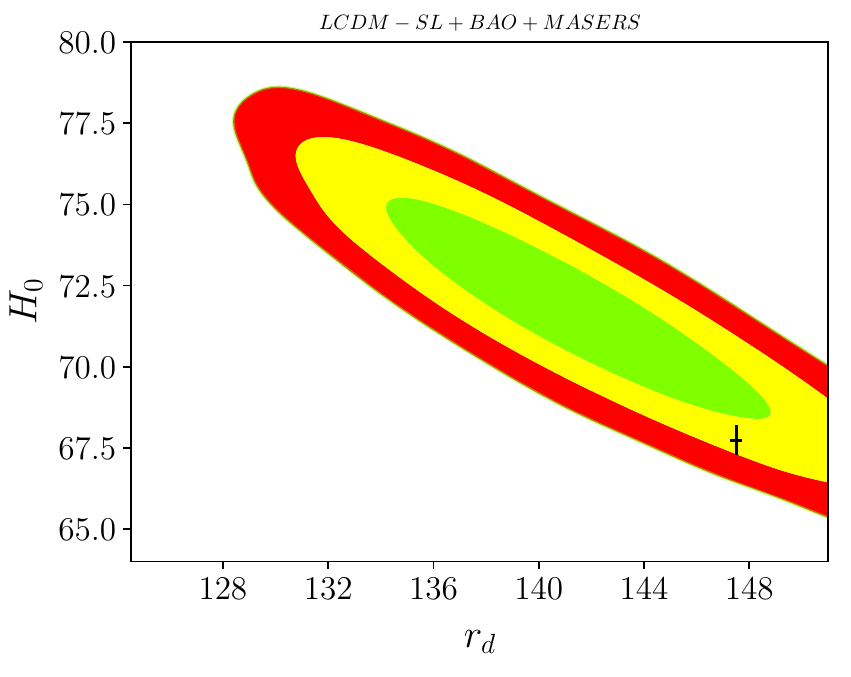}
\includegraphics[width=3.2in,height=2.2in]{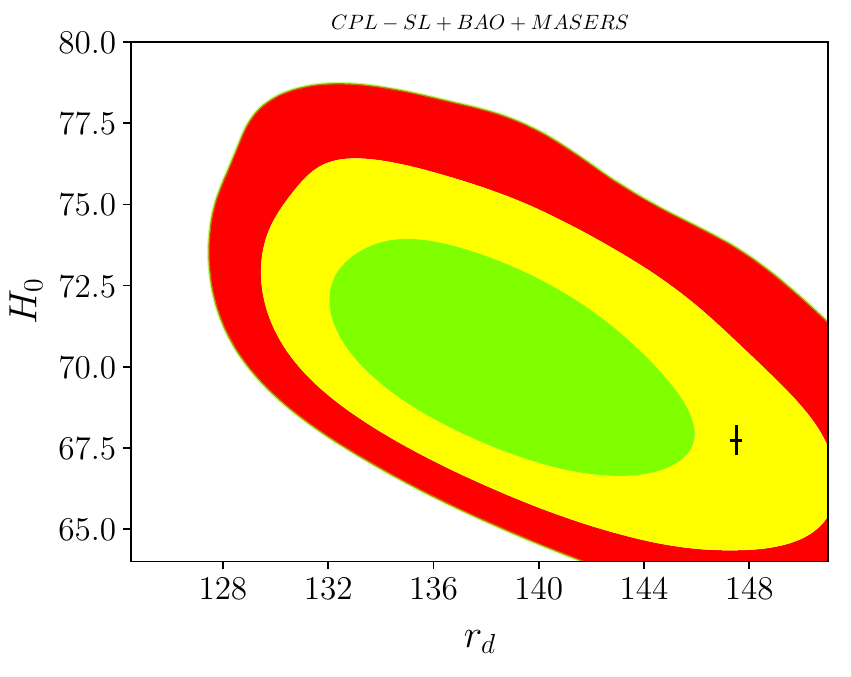}
\includegraphics[width=3.2in,height=2.2in]{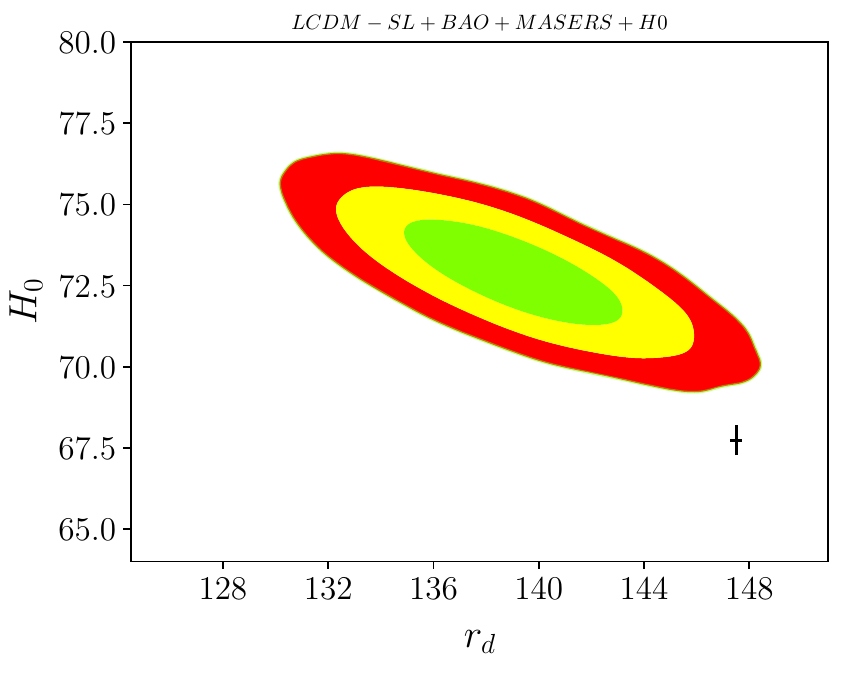}
\includegraphics[width=3.2in,height=2.2in]{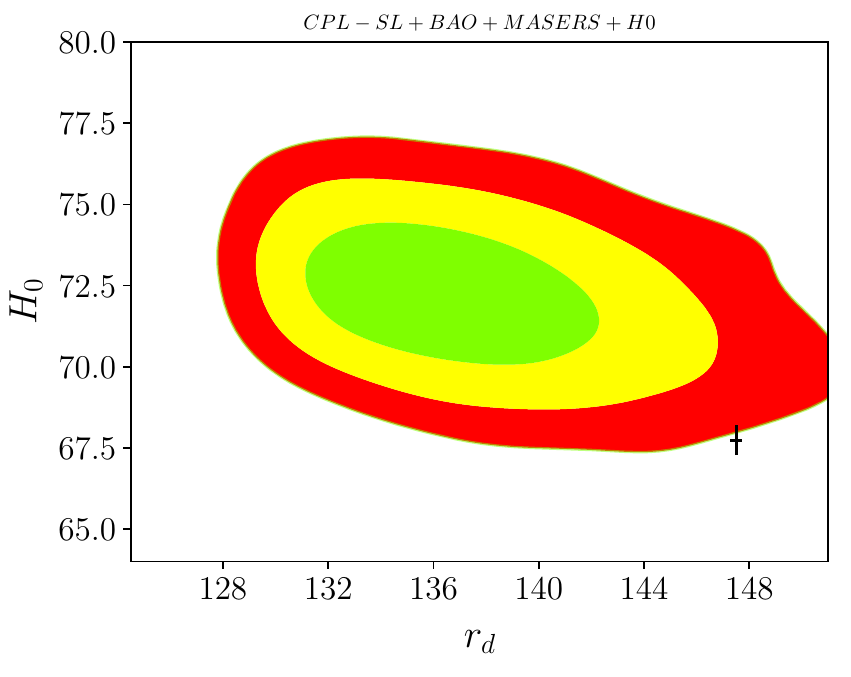}
\caption{An MCMC is used to calculate the standard error ellipses in $r_d$ and $H_0$ from the marginalized likelihood in the $\Lambda$CDM model (left) and CPL model (right) using BAO, lensing and masers (top) plus the distance ladder (bottom).   There is excellent agreement with the corresponding panels of Figs.~\ref{baolmpfig} and \ref{baolmhfig} which were calculated using the profile likelihoods.}
\label{mcmcfig}
\end{center}
\end{figure}

For brevity, here we provide only four examples of this comparison.  In Fig.~\ref{mcmcfig} one can see the two-dimensional standard error ellipses for $r_d$ and $H_0$ as calculated by the MCMC in the case of the $\Lambda$CDM models using BAO, strong lensing and megamasers with and without the distance ladder measurement of $H_0$.  Note that, since the priors are flat in the region plotted, the marginalized likelihood and the posterior probability distribution function are proportional in this region and so the confidence regions derived from them agree.

\section{Discussion}

\subsection{Consistency with CMB Data}

So far we have seen that BAO data together with strong lensing time delays and with or without masers prefers a value of $r_d$ which about 2$\sigma$ below the Planck benchmark value which is determined assuming $\Lambda$CDM.  Statistically significant tension arises only if one includes the local distance ladder measurement of $H_0$, which does little to shift the best fit value of $r_d$ but does reduce its uncertainty somewhat.  However if future BAO and lensing measurements confirm their best fit value with increasing precision, then $r_d$ indeed needs to be revised downwards.  There are only three ways in which this can happen.

\begin{figure} 
\begin{center}
\includegraphics{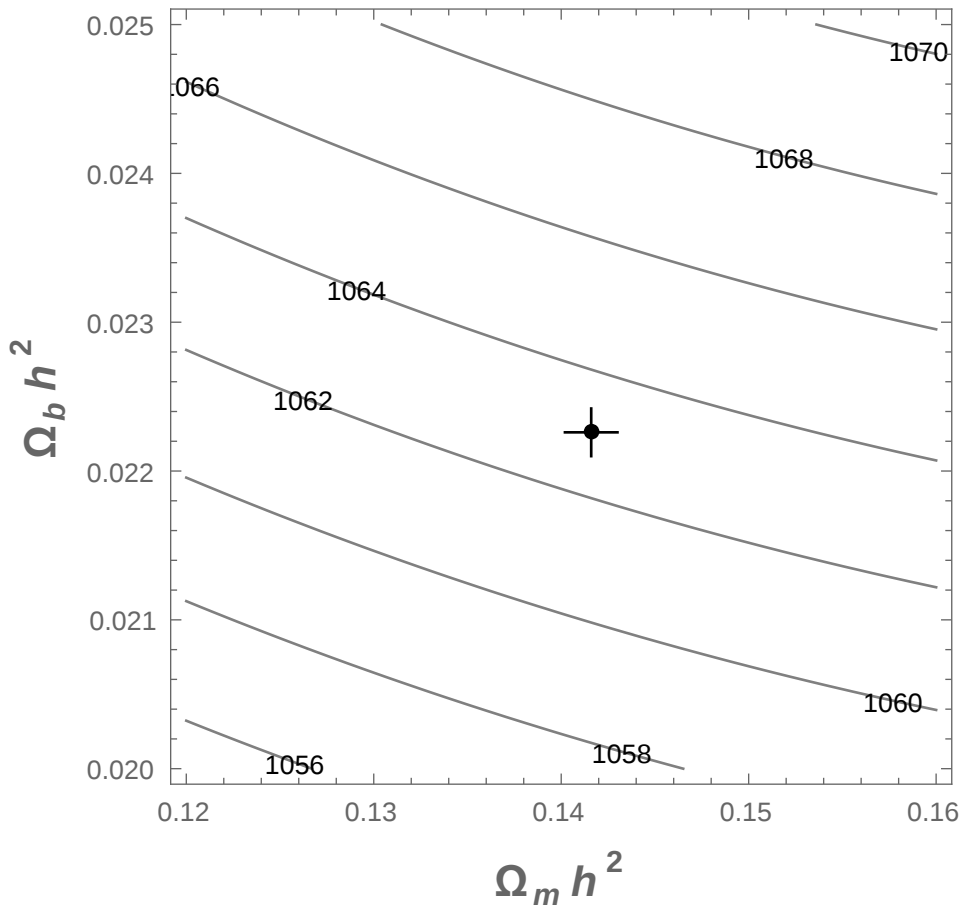}
\caption{The redshift of the drag epoch.}
\label{zfig}
\end{center}
\end{figure}

First, recall that $r_d$ here is measured in comoving coordinates.   Thus a smaller value of $r_d$ does not mean that the true, metric acoustic scale is smaller.  It may simply be at a higher redshift.  This would require an increase in the redshift $z_d$ to the drag epoch of between 50 and 100.  Due to the weak temperature dependence of the Saha equation together with the precisely determined CMB temperature, such a shift is actually remarkably difficult to achieve.  For example, in Fig.~\ref{zfig} we plot the redshift of the drag epoch according to the fitting formula of Ref.~\cite{hu96}, given a standard CDM cosmology with no dark radiation.  One may observe that such a large jump in redshift would require a dramatic change in the cosmological model, likely to be quite inconsistent with mass estimates from clusters, big bang nucleosynthesis (BBN), etc.  To regain consistency one would require a rather dramatic change in the model, and so we conclude that a simple reduction of $z_d$ is quite unlikely to resolve this tension.

The second possibility is that the speed of sound of the primordial plasma needs to be reduced.  This can be achieved by increasing the baryon to radiation ratio.  This ratio is constrained by BBN and also by the ratios of the even to the odd acoustic peaks.  One may attempt to loosen the former constraint with dark radiation or extremely early dark energy,  But what about the ratios of the even to the odd peaks?

Each cosmological model gives some transfer function from the primordial perturbations to the observed CMB power spectrum.  If the initial fluctuations are adiabatic, then they are characterized by a single function of the wavenumber.  Given the TT power spectrum and any model, and so any transfer function, there is always some adiabatic initial condition which leads to this spectrum.  Of course if one insists on a single field inflationary model then many such choices are excluded, but we are attempting to avoid such assumptions here.  Therefore, if one is free to choose adiabatic initial conditions as one pleases, one can fit any TT power spectrum.  Now it would require quite a fine-tuned initial spectrum to mimic the CMB acoustic peaks, since the peak locations would have to have coincidentally appeared in the initial conditions.  However broader features of the power spectrum could be mimicked by broad features of the initial conditions.  In the case of the damping scale it was shown in Ref.~\cite{spt} that this may be achieved with the standard logarithmic running of the spectral index of the primordial fluctuations.  What about even to odd peak ratios?  No simple logarithmic running would do this, however it would suffice to modify just the second and fourth peaks, and so the perturbation to the primordial fluctuation power spectrum need not be so fine-tuned.  We will examine this further in work in progress.

Finally, one may reduce $r_d$ by reducing the age of the Universe at the damping epoch.  This requires either a modification of gravity or else an increase in energy, which may be early dark energy, dark radiation or something more exotic like decaying dark matter.  It is the dark radiation proposal which has received the most attention in the literature.  In general, it is found that dark radiation is not sufficient to remove the tension, although it certainly plays a role in scenarios such as Refs.~\cite{12silk,12linder}.

Is this reduction of $r_d$ consistent with CMB constraints?  Perhaps one of the quantities best constrained by the CMB is the angular size of the acoustic scale $\theta_*$.  Ignoring the small difference between the drag epoch and recombination, this is $r_d/D_A(z_d)$.  A reduction in $r_d$ therefore requires a proportional reduction in $D_A(z_d)$.  However with fixed dark energy dynamics, $D_A(z_d)$ is inversely proportional to $H_0$.  Since the BAO constraint fixes $r_d H_0$, our reduction in $r_d$ is already proportional to the increase in $H_0$ and so this angular size is automatically held fixed by the BAO constraint.  This is fortunate, since a shift in the acoustic scale shifts all of the peaks in the $l$ direction, which would require a very contrived initial perturbation spectrum if one attempted to cancel its effect using initial perturbations.

On the other hand, there are other CMB constraints which are not automatically satisfied by a shift in $r_d$ with a compensating shift in $H_0$.  One of these is the damping scale, which as is reviewed in Ref.~\cite{spt} scales as the square root of the age of the Universe at recombination, unlike $r_d$ which scales linearly.  This means that any mechanism which simply shifts the age at recombination, be it dark radiation or dark energy, will necessarily affect the damping scale.  The observed angular size of the damping scale will increase, corresponding to less damping and so more power at high $l$.  Of course this can be compensated by a running spectral index which provides less initial power at high $l$ as in Ref.~\cite{spt} or by dark matter-neutrino interactions as in Ref.~\cite{dmneut} and in the present context \cite{dmneutb}.  Such interactions are in fact always present in WIMP models of dark matter.  Dark radiation will also affect the redshift of matter-radiation equality, which again affects the ratio of power at large and small $l$ and so in principle can be corrected with an appropriate shift to the running spectral index, although not a running of the same form as that which served for the damping scale.  

We note however that adiabatic initial conditions only allow one function of freedom in the determination of these perturbations and so they can fit one  observed power spectrum, for example TT.  Polarization measurements break the degeneracy between the transfer function and the initial perturbations.  Also in Ref.~\cite{riesstrouble} one sees that high $l$ polarization measurements ruin the compatibility of a dark radiation model with the data.  As the proliferation of parameters in fits continues, the role of polarization measurements in breaking this degeneracy will be essential.

Recently Ref.~\cite{des17} has shown that the low value of $H_0$ obtained by Planck assuming $\Lambda$CDM can be obtained with no use of CMB data.  Instead the authors used BBN data, assuming $\Lambda$CDM with no dark radiation and minimal neutrino masses, to calculate $r_d$.  Furthermore various Dark Energy Survey measurements were included, essentially replacing the pre-recombination power spectrum measurements of the CMB with measurements of matter clustering in the recent Universe.  The authors claim that such measurements are statistically independent of the Planck CMB measurement and so their consistency with the Planck value, together with moderate tension with other determinations, suggests that overall the various data sets are consistent.  

One may on the other hand argue that Planck and the Ref.~\cite{des17} analysis both rely heavily on the application of $\Lambda$CDM to early Universe cosmology, and that together they are statistically inconsistent with the combination of low redshift datasets provided by the local distance ladder, strong lensing time delays and the analysis presented in our study.  These low redshift datasets do not make any cosmological assumptions regarding the early Universe.   An inconsistency between these two sets of datasets and analyses (one with and one without assuming $\Lambda$CDM in the early Universe) therefore suggests a modification of early Universe cosmology.  From this point of view the results of \cite{des17} are nonetheless important because one learns that this modification must not only preserve agreement with the power spectrum of the primordial plasma as seen in the CMB, but also the matter power spectrum at lower redshifts as observed by DES.  Of course these two are closely related and so it seems plausible that one model could do both.  We will investigate this problem in the sequel.

\subsection{Concluding Remarks}

In Ref.~\cite{linderdemodel} the authors claimed that a model with modified dark energy resolves the tension between the local distance ladder $H_0$ and Planck.  It was suggested that in the future one should investigate whether this resolution survives datasets such as BAO.  We believe that the current paper answers the question posed there.  BAO demands that any shift in $H_0$ be accompanied by a shift in $r_d$ and so in pre-recombination cosmology.   Various combinations of strong lensing, megamasers and the local distance ladder suggest shifts of $H_0$ at various confidence levels, and we found that when including BAO these translate into shifts of $r_d$ at about the same confidence levels.

In particular, consistency between a BAO determination of $r_d H_0$ and the angular acoustic scale $r_r/D_A(z_r)$ in the CMB implies that the angular diameter distance to recombination shifts inversely proportionally to $H_0$, with a small error due to the difference between the angular acoustic scale at recombination and at the drag epoch and uncertainties in late time dark energy.  However the product $H_0D_A(z_r)$ is determined entirely by the dark energy content and dynamics, and so we arrive at the following claim.

\noindent
{\bf{Claim:}} {\it{Given a shift in $H_0$, the consistency of BAO and the CMB angular acoustic scale measurements is maintained if and only if}}
\beq
\frac{1+z_r}{c}H_0 D_A(z_r)=\int_0^{z_r}\frac{dz^\prime}{E(z^\prime)}= \int_0^{z_r}\frac{dz^\prime}{\sqrt{\Omega_m(1+z^\prime)^3+(1-\Omega_m)e^{3\int_0^{z^\prime}\frac{1+w(z^{\prime\prime})}{1+z^{\prime\prime}}dz^{\prime\prime}}}}
\eeq
{\it{remains approximately invariant.}}

This does not suggest dynamical dark energy, quite on the contrary it states that at least one number calculated from the dark energy dynamics, intuitively an average acceleration, must rest invariant\footnote{This is consistent with Refs.~\cite{nankai,ruchika} which found no evidence for dynamical dark energy.}.  It is the early time cosmology that needs to adapt to accommodate a new measurement of $H_0$ in the local Universe.

\section* {Acknowledgement}
\noindent
JE is supported by NSFC grant 11375201 and the CAS Key Research Program of Frontier Sciences grant QYZDY-SSW-SLH006.  JE also thanks the Recruitment Program of High-end Foreign Experts for support.  Ruchika thanks the Council of Scientific and Industrial Research (CSIR), Govt. of India for a Junior Research Fellowship.  Ruchika and AS thank the Institute of Modern Physics, CAS for hospitality at the beginning of this collaboration.


\end{document}